\documentclass{elsart}

\usepackage{amssymb}
\usepackage{graphicx}
\usepackage{amsmath}

\begin{document}

\begin{frontmatter}

\title{Magnetic Field induced Dimensional Crossover Phenomena in Cuprate
Superconductors and their Implications}

\author{T. Schneider} and \author{J. M. Singer\thanksref{AA}}

\address{Physikinstitut, Universit\"{a}t Z\"{u}rich, 
Winterthurerstr. 190, CH-8057 Z\"{u}rich, Switzerland}

\thanks[AA]{Corresponding author. Tel.: +41 1 635 4017 fax: +41 1 635
5704, e-mail: jms@physik.unizh.ch}

\begin{abstract}
We discuss the occurrence of
crossing points in the magnetization - temperature $(m,T$) plane
within the framework of critical phenomena. It is shown
that in a two-dimensional superconducting slab of thickness $d_{s}$  
$m_{z}\left(\delta \right)$ versus temperature $T$ curves measured in different 
fields $\mathbf{H} = H\left( 0,\sin \left( \delta \right) ,\cos \left(
\delta \right) \right)$ will cross at the critical temperature
$T_{c}$ of the slab. In contrast, in a 3D anisotropic
bulk superconductor the crossing point occurs in the plot 
$m_{z}\left(\delta \right) /H_{z}^{1/2}$ versus $T$. 
The experimental facts that
2D crossing point features have been observed in ceramics and in single
crystals for $\mathbf{H}$ close to $\mathbf{H} = H\left(
0,0,1\right)$, but not for $\mathbf{H} = H\left( 0,1,0\right)$, is
explained in terms of an angle-dependent crossover field separating the
regions where 2D or 3D thermal fluctuations dominate. The measured
2D-crossing point data are used to estimate one of the fundamental
parameters of cuprate superconductors, the minimum thickness of the slab 
$\left( d_{s}\right) $, which remains superconducting. 
Our estimates, based on
experimental 2D-crossing point data for single crystals, reveal that this
length adopts material dependent values. Therefore, experimental data
for $T_{c}$ and $\lambda _{\Vert }^{2}\left( T=0\right) $, plotted in terms
of $T_{c}$ versus $1/\lambda _{\Vert }^{2}\left( T=0\right) $ will not tend
to a straight line with universal slope as the underdoped limit is approached.
Implications for magnetic torque measurements are also worked out.
\end{abstract}

\begin{keyword}
High-T$_c$ cuprates, dimensional crossover, fluctuations, XY scaling
\end{keyword}
\end{frontmatter}

\section{Introduction}

It is well established that cuprate superconductors exhibit a pronounced
an\-iso\-tropy in their thermodynamic and transport properties. Within the
framework of \ a Ginzburg - Landau description, this anisotropy is
characterized by the effective mass  anisotropy
\begin{equation}
M_{x}\approx M_{y}=M_{\parallel }\ll M_{\perp }=M_{z},\ \gamma =\sqrt{%
M_{\perp }/M_{\parallel }}\gg 1
\end{equation}
where $M_{\Vert ,\bot }$ denotes the effective masses perpendicular and
parallel to the $\mathrm{CuO_{2}}$ layers, respectively. . Transport and
magnetic torque measurements (see e.g. \cite{SchneiderKeller} and refs.
therein) revealed that  $\gamma $ adopts -- even close to optimum doping --
rather large values and exhibits a strong doping dependence, i.e. $\gamma $
rises sharply by approaching the underdoped limit. Noting then that $\gamma
\rightarrow \infty $ represents the 2D limit, one concludes that the
materials will progressively exhibit quasi 2D properties by approaching the
underdoped limit, and that at $T=0$ an insulator to superconductor
transition occurs. In this limit, the materials can then be viewed as a
stack of independent superconducting slabs of thickness $d_{s}$. This feature
appears to be a generic property of  cuprate superconductors, and it implies
a dimensional crossover from 3D-XY to essentially 2D-XY behavior as
the underdoped limit is approached \cite{SchneiderKeller}. The scaling theory
of quantum critical phenomena \cite{EPL} predicts that for 2D-XY systems the
relation 
\begin{equation}
\lim_{x\rightarrow x_{u}}T_{c}(x)\lambda _{\parallel }^{2}(x,T=0)=\frac{\Phi
_{0}^{2}d_{s}}{Q_{0}16\pi ^{3}k_{B}},
\label{EQ3}
\end{equation}
is universal. $\lambda $ is the magnetic penetration depth, $\Phi_{0}$ 
the flux quantum, $d_{s}$ the thickness of the
superconducting slab (or sheet), $k_{B}$ the Boltzmann constant and $Q_{0}$
adopts an universality class specific value. $x$ is the doping variable
relative to the underdoped limit $x_{u}$, corresponding to a critical
endpoint where $T_{C}$ vanishes. This relation appears to explain the
experimentally observed trend, that for a variety of cuprate
superconductors $T_{C}$ plotted versus $1/\lambda _{\parallel }^{2}(x,T=0)$
seems to approach a universal line as $T_{C}$ decreases in the underdoped
regime  \cite{Uemura}. It was found, however, that several cuprates
containing $\mathrm{CuO}$ chains exhibit a reduced values 
$1/\lambda _{\parallel}^{2}(x,T=0)$ compared to the universal line 
\cite{Bernhard,Shengelaya}. 
In this context it should be recognized that the available experimental
data are measured quite far away from the critical endpoint where Eq. (\ref{EQ3}) 
applies. Furthermore, the existence of an universal line would require 
a material independent value of
the slab thickness $d_{s}$. Such an universal value of $d_{s}$ appears to be
rather unlikely due to the large variations in the chemistry within the unit cell.

We explore within the framework of critical phenomena the effect of an 
applied angular dependent magnetic field on the thermodynamic properties 
of cuprate superconductors, with special emphasis on the angular
dependence of magnetization and magnetic torque. It is shown that in a 2D 
superconducting slab of thickness $d_{s}$ $m_{z}\left( \delta \right)$
versus $T$ curves measured in different fields  
$\mathbf{H} = 
H\left( 0,\sin \left( \delta \right) ,\cos \left( \delta \right) \right)$,
will cross at the transition temperature $T_{c}$ of the slab,
while in an anisotropic bulk (3D)
superconductor the crossing point occurs in plots $m_{z}\left( \delta
\right) /H_{z}^{1/2}$ versus $T$. The experimental fact that 2D-crossing
point features have been observed in ceramics and in single crystals for 
$\mathbf{H}$ close to $\mathbf{H}\ =H\left( 0,0,1\right)$,
but not for $\mathbf{H}\ =H\left( 0,1,0\right)$, is explained in
terms of an angle-dependent crossover field separating the regions where
2D or 3D thermal fluctuations dominate. The measured 2D-crossing point
data are then used to estimate one of the fundamental parameters of cuprate
superconductors, the minimum thickness of the slab $\left( d_{s}\right) $
which remains superconducting. Our estimates, based on experimental
2D-crossing point data for single crystals reveal,  that $d_{s}$ adopts
indeed a material dependent value. As a consequence, experimental data for $%
T_{c}$ and $\lambda _{\Vert }^{2}\left( T=0\right) $, plotted in terms of $%
T_{c}$ versus $1/\lambda _{\Vert }^{2}\left( T=0\right) $ should not tend to
straight line with universal slope as the underdoped limit is approached.
Implications for magnetic torque measurements are also worked out.

\section{Angular dependence of the magnetization and crossing point phenomena}

The  appropriate scaling form of the singular part of the free energy
density for an anisotropic bulk superconductor in an applied magnetic field $%
\mathbf{H}=(0,H_{y},H_{z})$ is \cite{Fisher,Schneider} 
\begin{equation}
f_{s}=\frac{Q_{3}^{\pm }k_{B}T}{\xi _{a}^{\pm }\xi _{b}^{\pm }\xi _{c}^{\pm }%
}G_{3}^{\pm }\left( z\right) ,  \label{EQ5}
\end{equation}
with the scaling variable 
\begin{equation}
z={\frac{(\xi _{a}^{\pm })^{2}H}{\Phi _{0}}}\sqrt{{\frac{M_{a}}{M_{c}}}\sin
^{2}(\delta )+{\frac{M_{a}}{M_{b}}}\cos ^{2}(\delta )}.
\end{equation}
$G\left( z\right) $ is an universal scaling function. We obtain for the
magnetization 
\begin{equation}
m=-\frac{\partial f_{s}}{\partial H},
\end{equation}
\begin{equation}
m\left( \delta \right) = -\frac{Q_{3}^{\pm }k_{B}T}{\Phi _{0}^{3/2}}
H^{1/2}\left( \frac{M_{c}^{2}}{M_{a}M_{b}}\right) ^{1/4}
\left( \cos^{2}\left( \delta \right) +\frac{M_{b}}{M_{c}}
\sin ^{2}\left( \delta \right)
\right) \frac{dG_{3}^{\pm }\left( z\right) }{dz}\frac{1}{\sqrt{z}},
\label{EQ7}
\end{equation}
where we used 
\begin{equation}
\frac{\xi _{i}^{\pm }}{\xi _{j}^{\pm }}=\sqrt{\frac{M_{j}}{M_{i}}}.
\label{EQ8}
\end{equation}
For the $z$-component (perpendicular to the $\mathrm{CuO_{2}}$-layers) the
appropriate scaling form then reads  
\begin{equation}
\frac{m_{z}}{H_{z}^{1/2}}=-\frac{Q_{3}^{\pm }k_{B}T}{\Phi _{0}^{3/2}}\left( 
\frac{M_{c}^{2}}{M_{a}M_{b}}\right) ^{1/4}\frac{dG_{3}^{\pm }\left( z\right) 
}{dz}\frac{1}{\sqrt{z}},  \label{EQ6}
\end{equation}
\begin{equation*}
\quad z={\frac{(\xi _{a}^{\pm })^{2}H_{z}}{\Phi _{o}}}\sqrt{{\frac{M_{a}}{M_{b}}}}.
\end{equation*}

\begin{figure*}[h]
\centering
\includegraphics[width=6.5cm,angle=0.5]{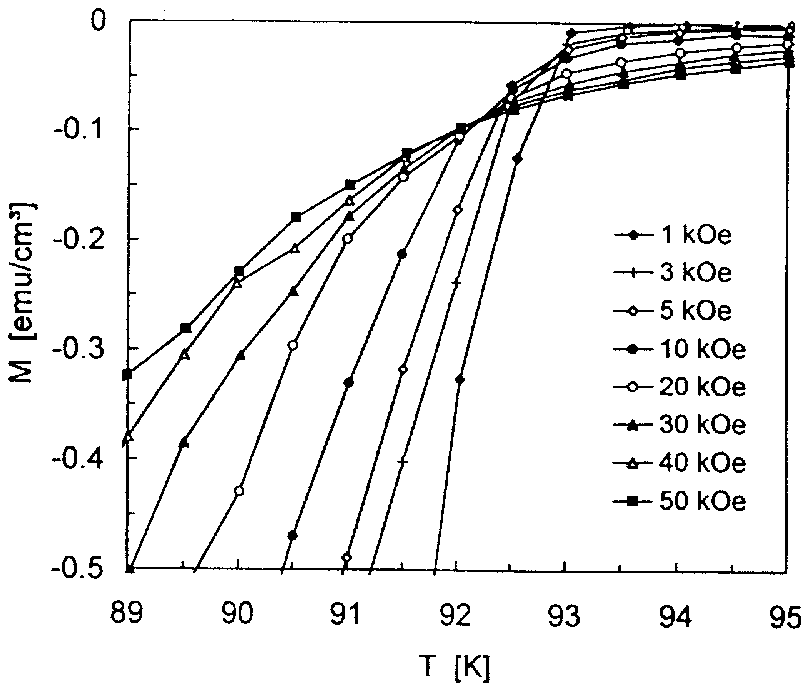}\includegraphics[width=6.5cm,angle=0.4]{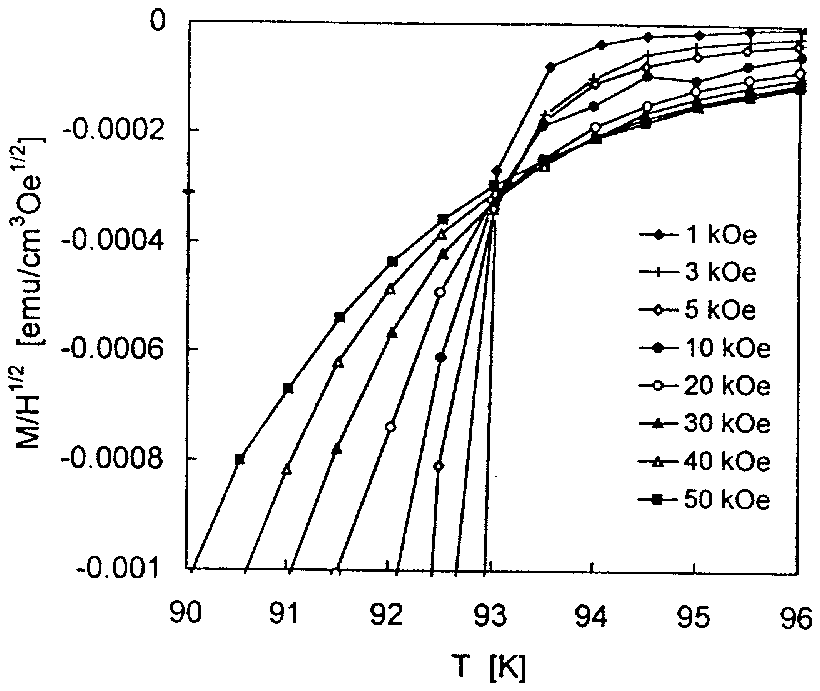}
\caption{$m_{z}$ versus $T$ (left panel) and $m_z/H_z^{1/2}$ versus $T$
(right panel) in various fields, $\mathbf{H}\parallel c$, for an optimally doped
$\mathrm{YBa_{2}Cu_{3}O_{7-\delta }}$ single crystal with $T_{c}=93K$ (Taken
from \cite{Junod}). }
\label{figjunod1}
\end{figure*}

In finite fields and close to $T_{c},$ where $\xi _{a}^{\pm }$ diverges
(i.e. $z\rightarrow \infty $), the existence of a magnetization obviously
requires 
\begin{equation}
\underset{z\rightarrow \infty }{\lim }\frac{dG_{3}^{\pm }\left( z\right) }{dz%
}\frac{1}{\sqrt{z}}=c_{3,\infty }^{\pm },  \label{EQ9}
\end{equation}
leading to 
\begin{equation}
\frac{m\left( T_{c},\delta \right) }{H^{1/2}} = -\frac{Q_{3}^{\pm }c_{3,\infty
}^{\pm }k_{B}T_{c}}{\Phi _{0}^{3/2}}
\left( \frac{M_{c}^{2}}{M_{a}M_{b}} \right) ^{1/4}
\left( \cos ^{2}\left( \delta \right) +\frac{M_{b}}{%
M_{c}}\sin ^{2}\left( \delta \right) \right) 
\label{EQ10}
\end{equation}
Thus, in a plot $m_{z}\left( \delta \right) /H_{z}^{1/2}$ versus $T$ the
data taken in different fields $H$ $\ $will cross at $T_{c}$. This behavior
has been well confirmed in $\mathrm{YBa_{2}Cu_{3}O_{7-\delta }}$ for $\delta
=0$ \cite{Hubbard,Junod}. As an example we show data from Junod et al. \cite
{Junod} in Fig. \ref{figjunod1}.

Using $\sqrt{M_{c}/M_{b}}\approx 8.95$, $\sqrt{M_{b}/M_{a}}\approx 0.83$ 
\cite{Schneider}, resulting from magnetic torque measurements, and ${%
m_{z}\left( T_{c}\right)}/{H_{z}^{1/2}}\approx 3\cdot 10^{-4} \mathrm{%
emu/cm^{3}Oe^{1/2}}$ (taken from Fig. \ref{figjunod1}) we obtain for the
universal number the estimate 
\begin{equation}
Q_{3}^{\pm }c_{3\infty }^{\pm }\approx 0.27,  \label{EQ11}
\end{equation}
which is close to the value $0.32$ obtained in the Gaussian approximation 
\cite{Prange}.

On the other hand,  considering a 2D-slab of thickness $d_{s}$, the scaling
form of the singular part of the free energy density (Eq.4) reduces to 
\begin{equation}
f_{s}=\frac{Q_{2}^{\pm }k_{B}T}{\xi _{a}^{\pm }\xi _{b}^{\pm }d_{s}}%
G_{2}^{\pm }\left( z\right) .  \label{EQ12}
\end{equation}
yielding for $M_{a}\approx M_{b}=M_{\Vert }$ 
\begin{eqnarray}
m_{2}\left( \delta \right)  &=&-\frac{Q_{2}^{\pm }k_{B}T}{\Phi _{0}d_{s}}%
\sqrt{\cos ^{2}\left( \delta \right) +\frac{1}{\gamma ^{2}}\sin ^{2}\left(
\delta \right) }\frac{\partial G_{2}^{\pm }\left( z\right) }{\partial z}, 
\notag \\
z &=&{\frac{(\xi _{a}^{\pm })^{2}H}{\Phi _{o}}}\sqrt{\cos ^{2}\left( \delta
\right) +\frac{1}{\gamma ^{2}}\sin ^{2}\left( \delta \right) }.  \label{EQ13}
\end{eqnarray}
\begin{figure}[h]
\centering
\includegraphics[width=6.5cm,angle=0.5]{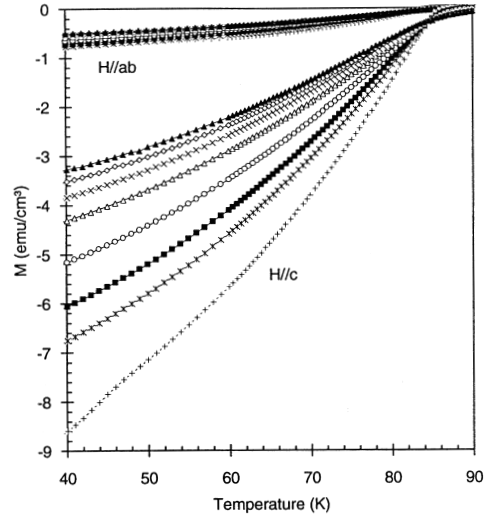}
\caption{$m$ versus $T$ for a
$\mathrm{Bi_{2}Sr_{2}CaCu_{2}O_{8}}$ single crystal with $T_c=85K$ in the
vicinity of $T_{c}$ (taken from \cite{Junod1}). Upper set
of curves: the field is parallel to the $\mathrm{CuO_2}$ planes. Lower set
of curves: the field is parallel to the $c$-axis. $H=1kOe$ (+) to $H=50kOe$
(filled triangles).}
\label{figjunod2}
\end{figure}
At the transition temperature of the slab, $T_{c}^{slab}$, $\xi _{a}^{\pm }$
diverges, i.e. in finite fields $z\rightarrow \infty $. Again, the
existence of a magnetization requires 
\begin{equation}
\underset{z\rightarrow \infty }{\lim }\frac{dG_{2}^{\pm }\left( z\right) }{dz%
}=c_{2,\infty }^{\pm },  \label{EQ14}
\end{equation}
and, thus, Eq.(\ref{EQ13}) can be rewritten as 
\begin{equation}
m_{2}\left( \delta ,T_{c}^{slab}\right) =-\frac{Q_{2}^{\pm }c_{2,\infty
}^{\pm }k_{B}T_{c}^{slab}}{\Phi _{0}d_{s}}\sqrt{\cos ^{2}\left( \delta
\right) +\frac{1}{\gamma ^{2}}\sin ^{2}\left( \delta \right) },  \label{EQ15}
\end{equation}
where $Q_{2}^{\pm }c_{2,\infty }^{\pm }$ is an universal number. As a
consequence, in a plot $m\left( \delta \right) $ versus $T$ for fixed $%
\delta $, the graphs taken in different fields will now cross at $%
T_{c}^{slab}$. To provide an estimate for $Q_{2}c_{2,\infty }$, we note that
the Gaussian approximation yields \cite{Klemm,Gerhardts} 
\begin{equation}
Q_{2}^{\pm }c_{2,\infty }^{\pm }=0.52.  \label{EQ16}
\end{equation}

\begin{figure*}[h]
\centering
\includegraphics[width=6.5cm]{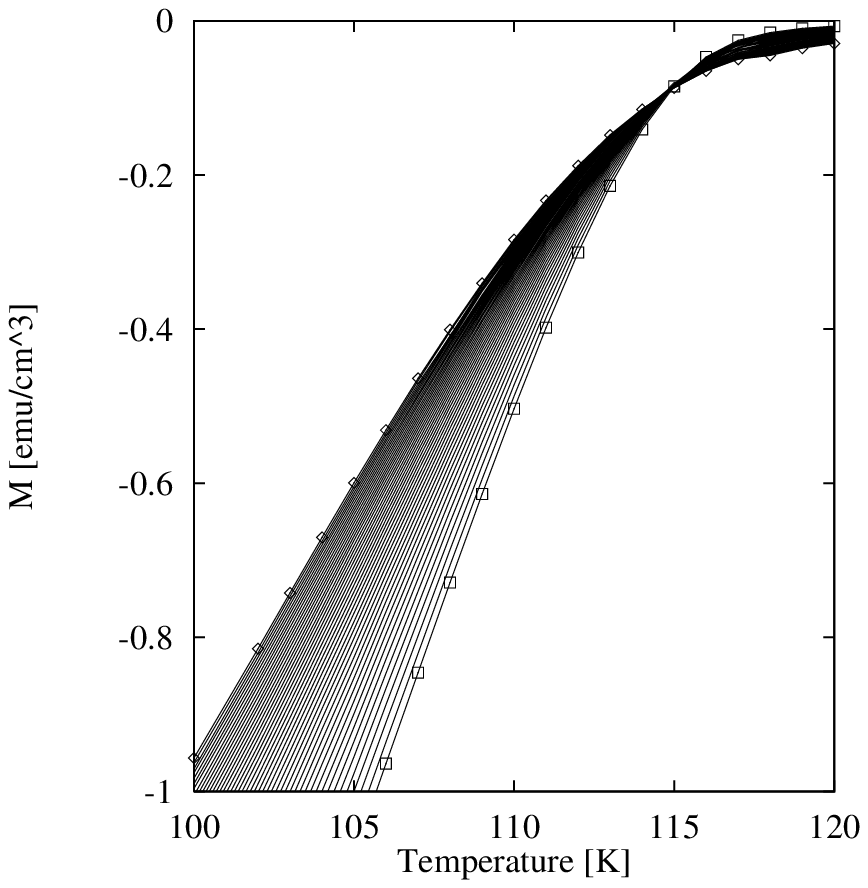}\includegraphics[width=6.5cm]{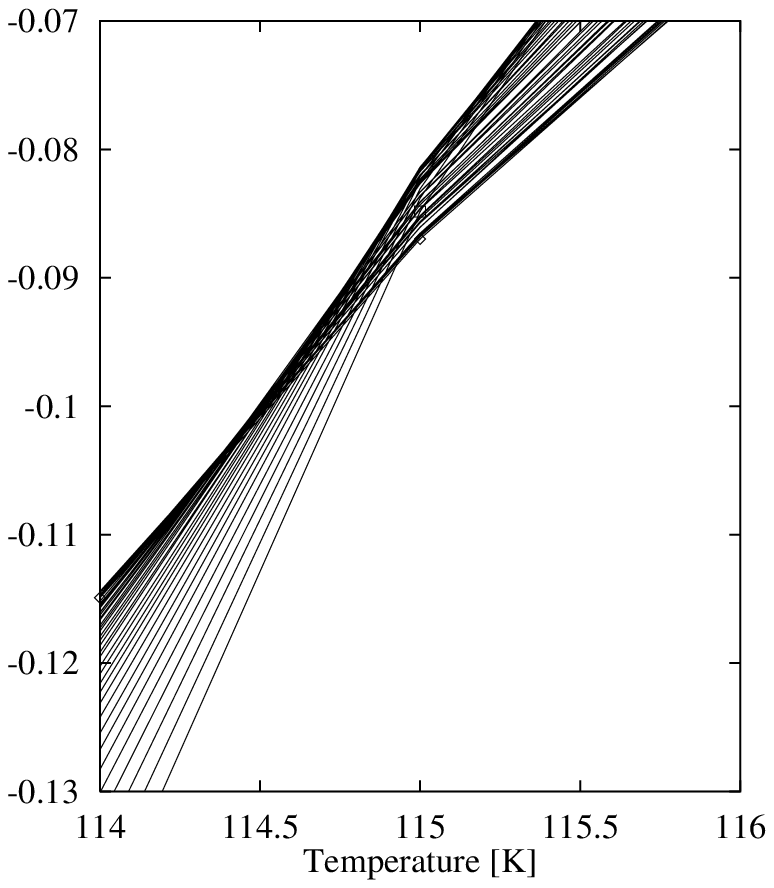}
\caption{Magnetization $m_{z}$ vs. $T$ for a $Tl-1223$ single crystal with $%
T_{c}=116K$ for various applied fields $H=10kOe$ (square) to $H=55kOe$
(diamond) perpendicular to the $\mathrm{CuO_2}$-layers.
The right panel shows a blow-up of the crossing region. We note
that although Tl-1224 is a quite anisotropic compound, the crossing point is
for lower fields rather a \textit{region} than a point in a strict sense.
Data courtesy G. Triscone \cite{Triscone}. }
\label{figtriscone}
\end{figure*}

In recent years considerable evidence for the occurrence of \ this
2D-crossing point phenomenon has been accumulated for sufficiently
anisotropic materials \ for ceramic samples and in single crystals for
fields perpendicular to the layers. 

\begin{figure}
\centering
\includegraphics[width=7cm]{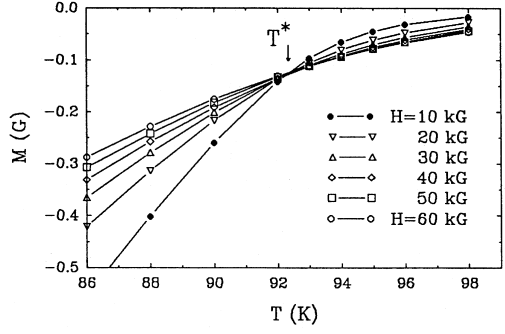}
\caption{$\bar{m}$ versus $T$ for randomly oriented polycrystalline $\mathrm{%
HgBa_{2}CuO_{4+\protect\delta }}$, applied field is perpendicular to the
$\mathrm{CuO_2}$-layers. (Taken from \cite{Thompson}. }
\label{figthompson}
\end{figure}

To illustrate this behavior we show in Figs. \ref{figjunod1} - \ref{figthompson} 
experimental data for an optimally doped $\mathrm{YBa_{2}Cu_{3}O_{7-\delta }}$
single crystal, a $\mathrm{Bi}_{2}\mathrm{Sr_{2}CaCu_{2}O_{8+\delta }}$ 
single crystal, a $\mathrm{TlBa_{2}Ca_{2}Cu_{3}O_{9}}$ single crystal
randomly oriented polycrystalline $\mathrm{HgBa_{2}CuO_{4+\delta }}$.
From Fig. \ref{figjunod1} (left panel) it is seen that in optimally doped 
$\mathrm{YBa_{2}Cu_{3}O_{7-\delta }}$, corresponding to the most isotropic material,
there is no well defined 2D-crossing point, while in the more anisotropic
materials,\\  $\mathrm{Bi}_{2}\mathrm{Sr_{2}CaCu_{2}O_{8+\delta }}$, 
$\mathrm{HgBa_{2}CuO_{4+\delta }}$ and $\mathrm{TlBa_{2}Ca_{2}Cu_{3}O_{9}}$
for $H\Vert c$ a reasonably well defined 2D-crossing point occurs over more 
than three decades of the field (see e.g. Fig. \ref{figthompson}), given
the limited resolution in both, magnetization and temperature. Indeed, even
in a highly anisotropic $\mathrm{TlBa_{2}Ca_{2}Cu_{3}O_{9}}$ single
crystal, the 2D-crossing point is seen to transform on a finer grid to a
crossing region (Fig. \ref{figtriscone}, right panel), revealing the presence of a finte
interslab coupling. Moreover, for $H\bot c$, the 2D-crossing point
phenomenon is not observed, even in 
$\mathrm{Bi_{2}Sr_{2}CaCu_{2}O_{8}}$, one of the most anisotropic cuprates \cite{Junod1}.

\begin{figure}[h]
\centering
\includegraphics[width=5.5cm]{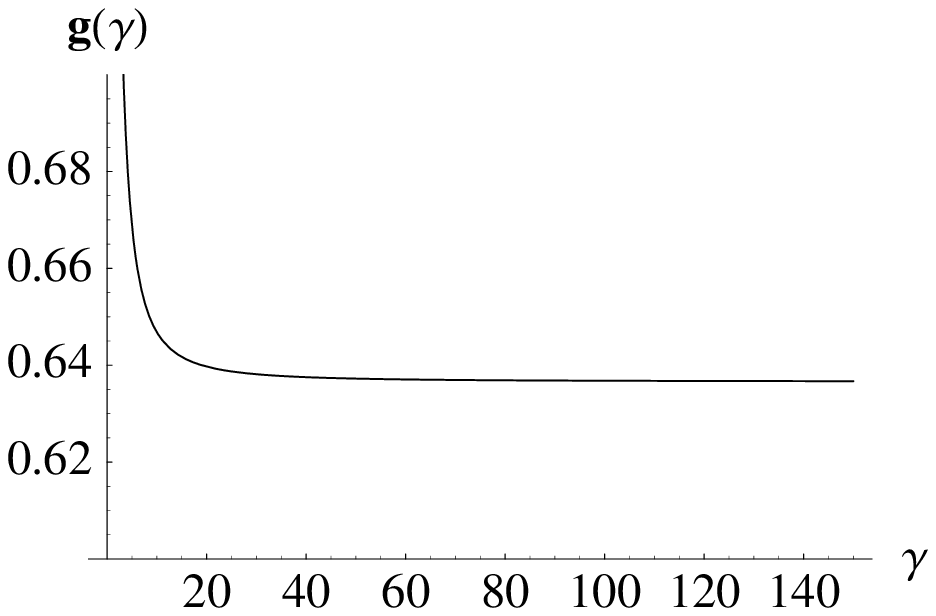}
\caption{$g(\gamma )$. }
\label{figgamma}
\end{figure}

Before we turn to this point and the angular dependence of the magnetization
in single crystals, it is useful to treat ceramic samples. Assuming randomly
oriented grains the crossing point relation (\ref{EQ15}) transforms into 
\begin{eqnarray}
\overline{m_{2}}\left( T_{c}^{slab}\right)  &=&-\frac{k_{B}T_{c}^{slab}}{%
\Phi _{0}d_{s}}g\left( \gamma \right) Q_{2}^{\pm }c_{2,\infty }^{\pm }, 
\notag \\
\ g\left( \gamma \right)  &=&\frac{1}{\pi }\int_{0}^{\pi }d\delta \sqrt{\cos
^{2}\left( \delta \right) +\frac{1}{\gamma ^{2}}\sin ^{2}\left( \delta
\right) }.  \label{EQ17}
\end{eqnarray}
From Fig. \ref{figgamma} it is seen that the variations of 
$g\left( \gamma \right)$, in the range of the relevant $\gamma $ values, is
rather small. 

\begin{figure}
\centering
\includegraphics[width=7.5cm]{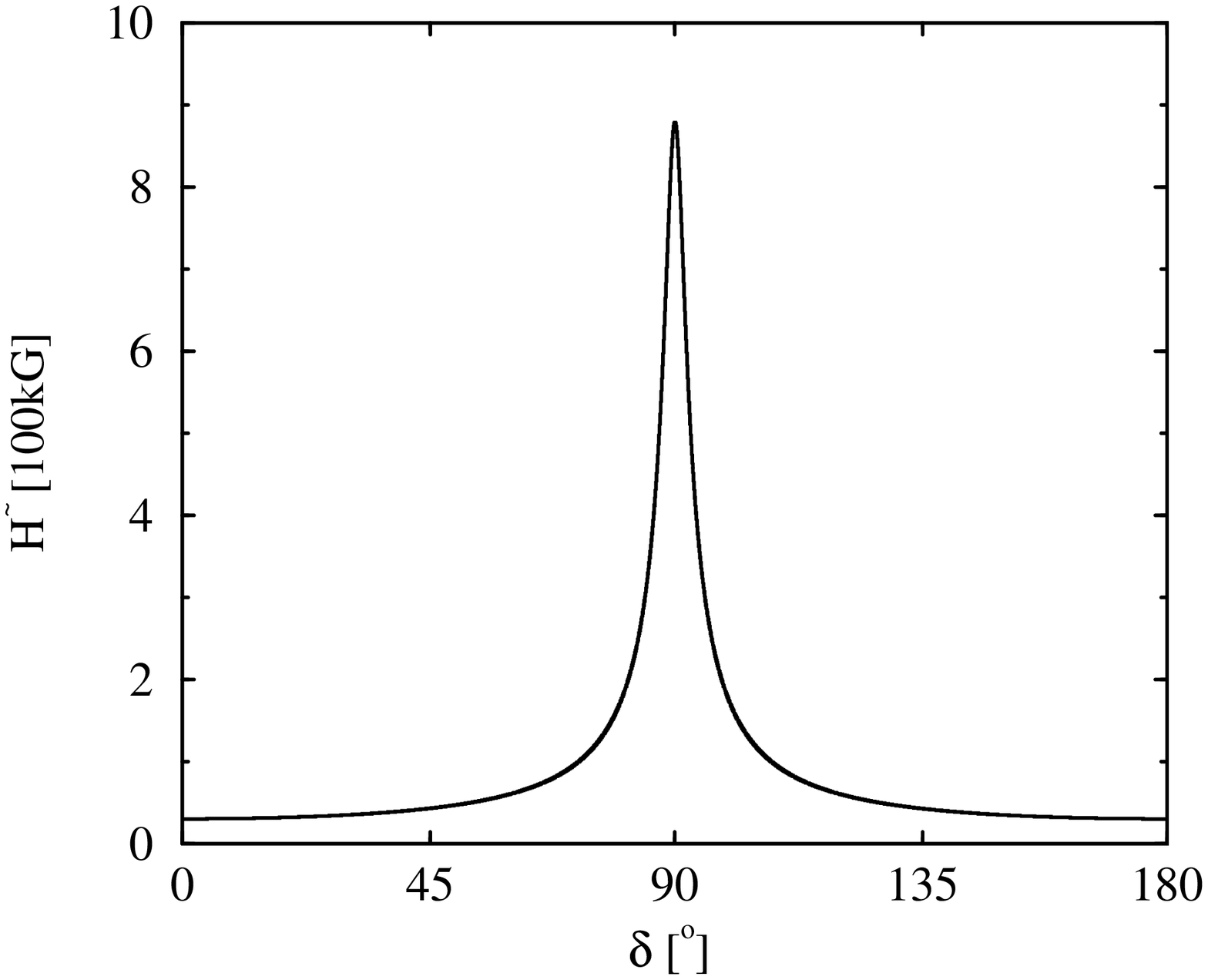}
\caption{$\tilde{H}\ [T]$ versus $\protect\delta $ for $\protect\xi _{\Vert
}\left( T^{\ast }\right) \approx 260.7\AA $ corresponding to $\mathrm{%
HgBa_{2}CuO_{4+\protect\delta }}$. }
\label{fighstern}
\end{figure}

The observation of the 2D crossing point features in ceramic samples and
single crystals with $H\Vert c$ seems to suggest a dimensional crossover,
which may be understood as following: approaching the bulk transition
temperature $T_{c}$ from below, $\xi _{c}^{-}$ diverges as 
\begin{equation}
\xi _{c}^{-}\left( T\right) =\xi _{c0}^{-}\left( 1-\frac{T}{T_{c}}\right)
^{-\nu },\nu \approx 2/3.  \label{EQ18}
\end{equation}
Lowering the temperature below $T_{c}$ , $\xi _{c}^{-}\left( T\right) $ can
become smaller than $s$, the crystallographic $\mathrm{CuO_{2}}$ layer
repeat distance. This causes a separation of the 3D bulk superconductor into
nearly uncoupled 2D - superconducting slabs of  thickness $d_{s}$, provided
that the anisotropy $\gamma $ is sufficiently large. This, in turn, renders
the critical amplitude $\xi _{c0}^{-}$ to be sufficiently small, so that 
\begin{equation}
\xi _{c}^{-}\left( T\right) <s  \label{EQ19}
\end{equation}
is satisfied for $T<T_{s}<T_{c}$ ($\xi _{c}^{-}\left( T_{s}\right) =s$). $%
T_{s}$ is the \textit{slab decoupling temperature}, i.e. the order parameter
fluctuations in adjacent slabs are only weakly correlated for $T<T_{s}$, and
the system corresponds to a stack of nearly independent quasi-2D
superconducting slabs of thickness $d_{s}$. In a strongly anisotropic system 
$T_{c}^{slab}$ is close to, but lower than the bulk transition $T_{c}$. $%
T_{c}^{slab}$ can be estimated from the bulk transition temperature in terms
of the finite size expression 
\begin{equation}
\xi _{c}^{-}\left( T_{c}^{slab}\right) =d_{s}\approx \xi _{c0}^{-}\left( 1-%
\frac{T_{c}^{slab}}{T_{c}}\right) ^{-\nu },\ \nu \approx 2/3.  \label{EQ20}
\end{equation}
Clearly (by definition), around $T_{c}^{slab}$ two-dimensional fluctuations
will dominate. Using for $\mathrm{YBa_{2}Cu_{3}O_{7-\delta }}$ ($T_{c}=91.5\
K$) $\xi _{c0}^{-}=1.64\AA $, $s=8.2\AA $ and for $\mathrm{%
HgBa_{2}CuO_{4+\delta }}$ ($T_{c}=95.6K$) $\xi _{c0}^{-}=0.94\AA $, $s=9.51%
\AA $ \cite{Schneider} as well as 
\begin{equation}
\xi _{c}^{-}\left( T_{s}\right) =s\approx \xi _{c0}^{-}\left( 1-\frac{T_{s}}{%
T_{c}}\right) ^{-\nu },  \label{EQ21}
\end{equation}
we find that $T_{s}\approx 83.3K$ for $\mathrm{YBa_{2}Cu_{3}O_{7-\delta }}$
and $T_{s}\approx 92.6K$ for $\mathrm{HgBa_{2}CuO_{4+\delta }}$. A condition
to observe in a bulk system a reasonably well defined 2D crossing point
phenomena is then 
\begin{equation}
T^{\ast }\approx T_{c}^{slab}\lessapprox T_{s}<T_{c},\ \ z\left( T^{\ast
}\right) \gg 1,  \label{EQ22}
\end{equation}
so that 2D critical fluctuations dominate around $T^{\ast }$ and the 2D
crossing point equations (\ref{EQ15}) and (\ref{EQ17}) apply (with $z\left(
T^{\ast }\right) \rightarrow \infty $). At $T_{s}$ a quasi-2D to 3D
crossover occurs, and around the bulk $T_{c}$ 3D-fluctuations determine the
critical behavior. In agreement with experiment (Fig. \ref{figjunod1}),
optimally doped $\mathrm{YBaCu_{3}O_{7-\delta }}$ is not expected to exhibit
2D crossing point features, because $T_{s}$ is much too far away from $T_{c}$.
In $\mathrm{HgBa_{2}CuO_{4+\delta }}$, however, condition (\ref{EQ22}) is
partially satisfied $\left( T_{s}\lessapprox T_{c}\right) $, but the
appearance of 2D crossing point features additionally requires $T^{\ast
}\approx T_{c}^{slab}\lessapprox T_{s}$, so that 2D fluctuations will
dominate. $\xi _{\Vert 0}^{-}$ and the scaling variable $z$ become so large
that Eq. (\ref{EQ15}) or (\ref{EQ17}) applies. In $\mathrm{%
HgBa_{2}CuO_{4+\delta }}$ the $T^{\ast }$ phenomenon has been observed in
randomly oriented polycrystalline samples for $10kG<H<60kG$ \cite{Thompson}
(see Fig. \ref{figthompson}). With the aid of Eq. (\ref{EQ13}) we introduce
an angular dependent ''crossover-field'' 
\begin{equation}
\frac{H^{\ast }\left( \delta \right) }{z}=\frac{\Phi _{0}}{\xi _{\Vert
}^{2}\left( T^{\ast }\right) }\frac{1}{\sqrt{\cos ^{2}\left( \delta \right) +%
\frac{1}{\gamma ^{2}}\sin ^{2}\left( \delta \right) }},  \label{EQ23}
\end{equation}
and, noting that a dominance of 2D fluctuations at $T^{\ast }$ requires $%
z\gg 1$ (Eq. (\ref{EQ22})), we obtain the condition 
\begin{equation}
H^{\ast }\left( \delta \right) >>\widetilde{H}\left( \delta \right) =\frac{%
\Phi _{0}}{\xi _{\Vert }^{2}\left( T^{\ast }\right) }\frac{1}{\sqrt{\cos
^{2}\left( \delta \right) +\frac{1}{\gamma ^{2}}\sin ^{2}\left( \delta
\right) }}  \label{EQ24}
\end{equation}
for the angular dependence of the field strength, where a well defined
crossing point phenomenon can be observed. In other words, 
$\tilde{H} \left( \delta \right)$ may also be viewed as a crossover field, separating
the regions where 2D- respectively 3D-fluctuations are essential. The angular
dependence of $\tilde{H}$ is shown in Fig. \ref{fighstern} for 
\begin{equation*}
\xi _{\Vert }\left( T^{\ast }\right) \approx \xi _{\Vert 0}\left( 1-T^{\ast
}/T_{c}\right) ^{-2/3}\approx 260.7A,
\end{equation*}
with $\xi _{\Vert 0}=\gamma \xi _{\bot 0}$ and the previously used
parameters for $\mathrm{HgBa_{2}CuO_{4+\delta }}$.

The angular dependence of $\widetilde{H}$, shown in Fig. \ref{fighstern}
reveals that for moderate field strength the observation of  2D-crossing
point features is restricted to ceramic samples or single crystals for
fields $H\Vert c$  ($\delta =0^{o})$, while for $\delta $ values around $%
90^{o}$ considerably higher fields are needed. Consequently, the occurrence
of a reasonably well defined crossing phenomenon in ceramics or randomly
oriented polycrystalline samples does not imply that 2D fluctuations
dominate per se, but most likely around $\delta =0$ only. Moreover, it
should be recognized that even though 2D fluctuations may dominate around $%
\delta =0$, the weak order parameter fluctuations along the $c$-direction
render to smear out the crossing point to a crossing region (as shown in
Fig. \ref{figtriscone}), its extend will decrease with increasing $\gamma $
(which, in turn, corresponds to decreasing interlayer coupling). In any
case, as long as $\gamma $ is finite, there will be, strictly speaking, no
2D-crossing point, because sufficiently close to the bulk $T_{c}$
3D-fluctuations dominate. In practice, however, the homogeneity of the
samples and the experimenal resolution are limited. For this reason the
observation of \ reasonably well defined 2D-crossing point features in
moderate fields is restricted to ceramics and single crystals with $H\Vert c$,
provided that $\gamma $ is sufficiently large.

Nevertheless, the observation of crossing point features offers the
possibility to estimate the slab thickness with the aid of Eq. (\ref{EQ15}) 
or (\ref{EQ17}),
provided that experimental values for $m_{2}\left( \delta =0,T_{c}^{slab}\right)
/T_{c}^{slab}$ or $\overline{m_{2}}\left( T_{c}^{slab}\right) /T_{c}^{slab}$
are available. Because the grains in ceramic samples are not necessarily randomly
oriented, we listed in Table \ref{Tabelle1}
some single crystal estimates. The quoted 
$d_{s}$-values have been derived from Eq. (\ref{EQ15}), using the Gaussian estimate
for $Q_{2}^{\pm }c_{2,\infty }^{\pm }$ (Eq. (\ref{EQ16})).

\begin{table}
\caption{$d_s$-values for various cuprate compounds.\label{Tabelle1}}
\begin{tabular}{|l||c|c|c|c|c|}
\hline
compound & $\underset{[K]}{T_{c}^{slab}}$ & $\underset{[emu/cm^{3}K]}
{m_{2}\left(\delta =0,T_{c}^{slab}\right) /T_{c}^{slab}}$ 
& $\underset{[\AA]}{d_{s}}$ & $\gamma $ & source \\ \hline
$\mathrm{HgBa_{2}Ca_{2}Cu_{3}O_{8+\delta }}$ & $127$ & $1.81\cdot 10^{-3}$ & $29$ & $%
52$ & \cite{tabref1} \\ \hline
$\mathrm{Bi_{2}Sr_{2}CaCu_{2}O_{8+\delta }}$ & $83.1$ & $3.54\cdot 10^{-3}$ & $10$ & $%
250$ & \cite{tabref2} \\ \hline
$\mathrm{Bi_{2}Sr_{2}CaCu_{2}O_{8+\delta }}$ & $84.5$ & $3.55\cdot 10^{-3}$ & $10$ & $%
250$ & \cite{Junod1} \\ \hline
$\mathrm{Bi_{2}Sr_{2}Ca_{2}Cu_{3}O_{10+\delta }}$ & $105$ & $1.14\cdot 10^{-3}$ & $30$
& $70$ & \cite{tabref4} \\ \hline
$\mathrm{TlBa_{2}Ca_{2}Cu_{3}O_{9+\delta }}$ & $115$ & $0.09\cdot 10^{-3}$ & $44$ & 
& \cite{Triscone} \\ \hline
$\mathrm{Tl_{2}Ba_{2}CuO_{6+\delta }}$ & $84.2$ & $3.80\cdot 10^{-3}$ & $9.12$ & $117$
& \cite{tabref6} \\ \hline
$\mathrm{Bi_{1.95}Sr_{1.65}La_{0.4}CuO_{6+\delta}}$ & $11.2$ & $2.41\cdot 10^{-3}$ & $14.4$ & & \cite{tabref7} \\ \hline
\end{tabular}
\end{table}

Noting that the quoted $d_{s}$-values depend on 
$Q_{2}^{\pm }c_{2,\infty }^{\pm }$, 
while their ratios are independent, the experimental data
confirms the expectation according to which $d_{s}$ adopts a material
dependent value. This implies, combined with the universal relation (\ref{EQ3}),
that experimental data for $T_{c}$ and $\lambda _{\Vert }^{2}\left(
T=0\right) $, plotted in terms of $T_{c}$ versus $1/\lambda _{\Vert
}^{2}\left( T=0\right) $ do not tend to straight line with universal slope
as the underdoped limit is approached.

\section{Magnetic torque}

Next we discuss implications of two-dimensional crossing point features for the
interpretation of the angular dependence of the magnetic torque. With this
technique, rather accurate and extended experimental data has been
accumulated, covering the full variation of $\delta $.

\begin{figure}
\centering
\includegraphics[width=7cm]{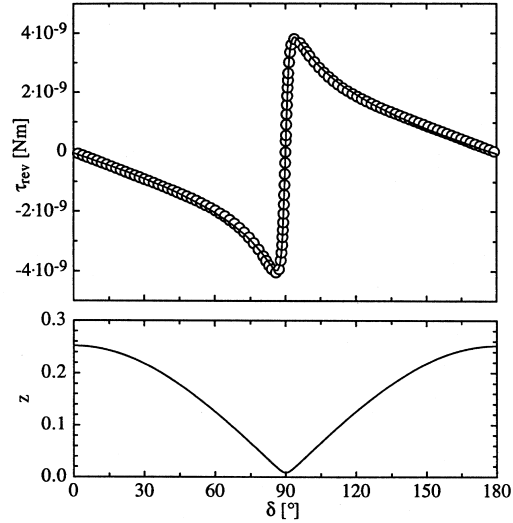}
\caption{a) $\protect\tau _{x}$\ versus $\protect\delta $ \ for a $\mathrm{%
HgBa_{2}CuO_{4+\protect\delta }}$ single crystal with $T_{c}=95.6$ $K$ at $%
H=1.4T$ and $T=90.9K$. Data points have been omitted for clarity. The solid
line is a fit to Eq. (\ref{EQ29}). b) $z$ versus $\protect\delta$. Taken
from \protect\cite{Hofer}. }
\label{fighofer}
\end{figure}

As an example we consider $\mathrm{HgBa_{2}CuO_{4+\delta }}$, where
according to Fig. \ref{fighstern} the dominance of 2D fluctuations around 
$\delta =90^{o}$ requires that $H\gg 8\cdot 10^{5}G$. Thus, for moderate
field strength and $T^{\ast }\lessapprox T_{c}$ anisotropic
three-dimensional critical fluctuations are expected to dominate around $%
\delta =90^{0}$. In this case the magnetic torque can be derived from
singular part of the 3D free energy density, Eq. (\ref{EQ5}), yielding in
the limit $z\rightarrow 0$, 
\begin{equation}
\mathcal{\tau }_{x}\left( \delta \right) =
{\frac{Q_{3}^{-}\Phi_{0}C_{3,0}^{-}H}{16\pi ^{3}\lambda _{\Vert 0}^{2}}}
\left( {1}-\frac{1}{\gamma }\right) 
{\frac{\cos (\delta )\sin (\delta )}{\sqrt{\cos
^{2}(\delta )+{\frac{1}{\gamma ^{2}}}\sin ^{2}(\delta )}}}\ln (z).
\label{EQ29}
\end{equation}
Since the torque signal vanishes at $\delta =0^{o}$, where the crossing
point features are known to occur (see Fig. \ref{figthompson}), this
technique turns out to be rather insensitive in the regime where 2D
fluctuations may dominate. To illustrate this fact we show in Fig. 
\ref{fighofer}a the measured angular dependence of the magnetic torque for a 
$\mathrm{HgBa_{2}CuO_{4+\delta }}$ single crystal with $T_{c}=95.6K$ at 
$H=1.4T$ and $T=90.9K$. The solid line is a fit to Eq. (\ref{EQ29}), yielding 
$\xi _{\Vert 0}^{-}\approx \gamma \xi _{\bot 0}^{-}\approx 27\AA $, $\gamma
\approx 28.8$ (as obtained in \cite{Hofer}) and $Q_{3}^{-}C_{3,0}^{-}\approx
0.7$ for the universal number. The corresponding $\delta $-dependence of $z$,
shown in Fig. \ref{fighofer}b, confirms that the data are in the
appropriate range ($z\rightarrow 0$), where Eq. (\ref{EQ29}) is applicable.

\begin{figure}
\centering
\includegraphics[width=7cm]{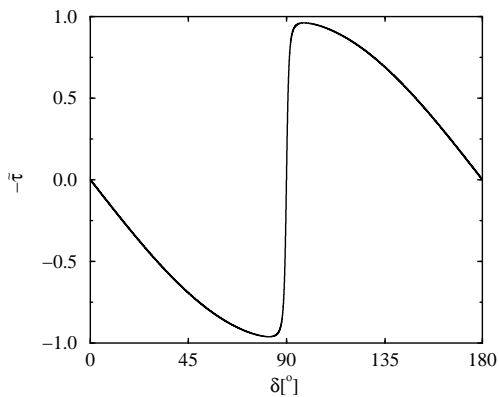}
\caption{$\tilde\protect\tau_x$ versus $\protect\delta$ close to $%
T_{c}^{slab}$ for $\protect\gamma =50$, according to Eq.(30). }
\label{figtorque2d}
\end{figure}

The remarkable fit must be attributed to the aforementioned facts: For the
field considered here ($H=1.4T$) and for the angular regime $%
\left(\delta\approx 90^o\right)$, where the magnetic torque provides a
strong signal, anisotropic 3D fluctuations dominate even close to $T^{\ast}$%
. Unfortunately, in the angular regime $\delta=0^o$, where around $T^{\ast }$
2D fluctuations are essential, the torque signal vanishes. To substantiate
this conclusion it is instructive to consider the angular dependence of the
magnetic torque for a slab of thickness $d_{s}$. Close to $T_{c}^{slab},$
where Eq. (\ref{EQ14}) applies, we obtain from the scaling form of the free
energy density with $m_{i}=-\partial f_{S}/\partial H_{i}$, $\tau
_{x}=m_{y}H_{z}-m_{z}H_{y}$ and $M_{x}\approx M_{y}$ 
\begin{equation}
\frac{\tau _{x}}{H}=-\frac{k_{B}TQ_{2}^{\pm }c_{2,\infty }^{\pm }}{\Phi
_{0}d_{s}}\left( 1-\frac{1}{\gamma }\right) \frac{\sin \left( \delta \right)
\cos \left( \delta \right) }{\sqrt{\cos ^{2}\left( \delta \right) +\frac{1}{%
\gamma ^{2}}\sin ^{2}\left( \delta \right) }}.  \label{EQ30}
\end{equation}
The resulting angular dependence is shown in Fig. \ref{figtorque2d} for $%
\gamma=50$. Comparing with the measured behavior in Fig. \ref{fighofer}a,
the pronounced deviations around $\delta=90^o$ fully confirm our
expectations that in the regime where the data of Fig. \ref{fighofer}a were
taken, i.e. for $\delta$ values where a appreciable torque signal appears,
anisotropic 3D fluctuations dominate.

We are then led to the important conclusion, that around $T^{\ast }$ the
magnetic torque probes, for moderate applied fields and $\delta \approx
90^{o}$, predominantly an\-iso\-tropic 3D fluctuations. To substantiate this
conclusion further, we note again that even in the highly anisotropic
compound , $\mathrm{Bi}_{2}\mathrm{Sr_{2}CaCu_{2}O_{8+\delta }}$ the
2D-crossing point phenomenon was not observed for $H\bot c$ \cite{Junod1}.

\section{Summary and Conclusions}

To summarize we have shown that an applied magnetic field, \\
$\mathbf{H}\ =H(0,\sin(\delta),\cos(\delta))$
 may induce a quasi 2D to 3D crossover leading to crossing points in the
magnetization - temperature $(m,T)$ plane. In a 2D-superconducting slab of
thickness $d_{s}$, the curves $m_{z}\left( \delta \right) $ versus $T$ 
taken in different fields will cross at $T_{c},$ while in an
anisotropic bulk (3D) superconductor the crossing point occurs in the plot 
$m_{z}\left( \delta \right) /H_{z}^{1/2}$ versus $T$. The experimental fact
that 2D-crossing point features have been observed in ceramics and in single
crystals for $\mathbf{H}$ close to $\mathbf{H}\ =H\left(0,0,1\right)$
 but not for $\mathbf{H}\ =H\left( 0,1,0\right)$ is
explained in terms of a $\delta$-dependent crossover field separating the
regions where 2D respectively 3D thermal fluctuations dominate. The measured
2D-crossing point data was then used to estimate one of the fundamental
parameters of cuprate superconductors, the minimum thickness of the slab 
$\left( d_{s}\right)$ which remains superconducting. Our estimates, based on
experimental 2D-crossing point data, reveal -- as expected -- that this length
adopts a material dependent value. As a consequence, experimental data for 
$T_{c}$ and $\lambda _{\Vert }^{2}\left( T=0\right) $, plotted in terms of 
$T_{c}$ versus $1/\lambda _{\Vert }^{2}\left( T=0\right) $ do not tend to a
straight line with universal slope as the underdoped limit is
approached. Finally, we have shown that angular dependent magnetic torque
measurements probe for moderate field strength and $\delta \approx 90^{o}$
predominantly anisotropic 3D fluctuations.

\begin{ack}

We benefitted from discussions with J. Hofer, H. Keller, M. Willemin, G.
Triscone, A. Junod, P. Martinoli and H. Beck. Part of the work was
supported by the Swiss National Science Foundation.

\end{ack}

\end{document}